\documentclass[%
 aip,
 amsmath,amssymb,
 reprint,%
]{revtex4-2}

\usepackage{graphicx}
\usepackage{dcolumn}
\usepackage{bm}
\usepackage[utf8]{inputenc}
\usepackage[T1]{fontenc}
\usepackage{mathptmx}

\begin{document}


\title{Temperature Dependence of Spin Pumping in Ni\textsubscript{81}Fe\textsubscript{19}/NbN Bilayer Thin Films }

\author{Sumesh Karuvanveettil} \affiliation{International Research Centre MagTop, Institute of Physics, Polish Academy of Sciences, PL-02668 Warsaw, Poland}
\author{Arathi  Moosarikandy} \affiliation{International Research Centre MagTop, Institute of Physics, Polish Academy of Sciences, PL-02668 Warsaw, Poland}
\author{Micha\l{} Chojnacki } \affiliation{International Research Centre MagTop, Institute of Physics, Polish Academy of Sciences, PL-02668 Warsaw, Poland} \affiliation{ Institute of Physics, Polish Academy of Sciences, PL-02668 Warsaw, Poland}
\author{ Krzysztof Fronc} \affiliation{International Research Centre MagTop, Institute of Physics, Polish Academy of Sciences, PL-02668 Warsaw, Poland}   
\author{Roman Minikayev} \affiliation{ Institute of Physics, Polish Academy of Sciences, PL-02668 Warsaw, Poland}
\author{Vinayak Shantaram Bhat} \email[]{Authors to whom correspondence should be addressed: Vinayak Bhat, vbhat@magtop.ifpan.edu.pl}  \affiliation{International Research Centre MagTop, Institute of Physics, Polish Academy of Sciences, PL-02668 Warsaw, Poland}
\vskip 0.25cm
\date{\today}

\date{\today}

\begin{abstract}
We present a comprehensive study of broadband spin pumping utilizing the inverse spin Hall effect phenomena in bilayer samples comprising Ni\textsubscript{81}Fe\textsubscript{19} (15 nm) and NbN (with NbN thickness varying from 20 nm to 140 nm), conducted over a temperature and frequency range spanning from 300 K to 4 K and 2 GHz to 12 GHz, respectively. Our investigations reveal  a systematic shift in ferromagnetic resonance fields, amplitude, and line widths as functions of both frequency and temperature. Notably, we observed a temperature-dependent increase in the spin Hall angle value, surpassing previously reported values. Furthermore, our results demonstrate a pronounced temperature dependence in the inverse spin Hall effect voltage, exhibiting a significant reduction below the $T_{c}$. This reduction in inverse spin Hall effect voltage  is accompanied by an increase in the linewidth of the ferromagnetic resonance mode.
\end{abstract}

\maketitle

Magnons (quanta of spin waves) are collective excitations within magnetic materials, enabling information processing without the need for charge transfer \cite{chumak2015magnon,kruglyak2010magnonics}. An essential aspect of magnonic devices involves the conversion of magnon signals into electrical signals. One effective method for achieving this electric detection of magnon spin currents is through the use of spin pumping via the inverse spin Hall effect (ISHE) phenomenon \cite{valenzuela2006direct,haldar2021functional,kimura2007room}.\\
Recent advancements have emerged on the controlled manipulation of  the interface between superconducting  and magnonic materials, leading to novel computing logic and memory paradigms. These innovations leverage not only the charge and spin properties of electrons, but also tap into the intriguing quantum attributes inherent in superconductivity. Significant strides have been taken in generating pure spin currents in superconducting Nb-interfaced hybrid films, where the efficiency of pure spin supercurrent has been successfully tuned. This breakthrough has given rise to a prototype superconducting spin-wave device, where the opening or closing of a pure spin supercurrent channel in Nb enables gateable spin-wave propagation \cite{jeon2020tunable}.\\
NbN, a typical s-wave type-II superconductor, possesses a bulk superconducting transition temperature ($T_c$) of approximately 16.5 K\textemdash higher than that for Nb\textemdash, a superconducting energy gap of 2.5 meV, a superconducting coherence length of 5 nm \cite{Linder2015}, and a large penetration depth of about 200 nm \cite{chockalingam2008superconducting}. However, research on ISHE in NbN is still in its early stages \cite{rogdakis2019spin}. Rogdakis et al. reported a comprehensive study of Y\textsubscript{3}Fe\textsubscript{5}O\textsubscript{12} (YIG)/NbN spin pumping, where the voltage signal disappeared below 50 K (a value significantly exceeding the $T_c$ of NbN). At low temperatures, YIG exhibits a notable increase in spin-wave damping, potentially attributed to slowly relaxing impurities during growth \cite{jermain2017increased}. Moreover, the challenges associated with on-chip integration and miniaturization for YIG arise due to critical conditions in deposition and fabrication \cite{li2019strong}. Therefore, it is crucial to explore ISHE characteristics in NbN hybridized with a well-established magnetic material, e.g., Ni\textsubscript{81}Fe\textsubscript{19} (Py), in close proximity to the $T_c$ of NbN. Furthermore, NbN has gained prominence as a superconducting microwave resonator for magnon-photon coupling, owing to its higher $T_c$ \cite{li2019strong}. Hence, investigating the effects of interfacing Py with NbN near its $T_c$ becomes imperative.\\
Here, we present the results of broadband spin pumping experiments [Fig. \ref{Fig1}(a)] performed on Py/NbN bilayers within a temperature  range spanning from 300 to 4 K and covering frequencies from 2 to 12 GHz. The Py/NbN bilayers were  grown using room temperature magnetron sputtering, with an argon-nitrogen ratio of 10:1 during the NbN deposition, at  a pressure of 10\textsuperscript{-3} mbar after achieving a base pressure of 10\textsuperscript{-9} mbar. We deposited five different thicknesses of NbN in-situ on top of a 15 nm thick Py layer: samples denoted as S-1, S-2, S-3, S-4, and S-5, with approximate NbN  thicknesses of 20, 40, 60, 80, and 140 nm, respectively. All of these NbN thicknesses fall below the typical NbN penetration depth of 200 nm \cite{chockalingam2008superconducting}. A reference sample (S-0) consisting of a 15 nm thick Py layer capped with a 3 nm thick layer of Al to prevent oxidation was also prepared.  We determined the thickness-dependent $T_{c}$ in these samples through the four-probe resistance technique [Fig. \ref{Fig1}(b)]. Our observations revealed the lowest $T_{c}$ in sample S-1, featuring a 20 nm thick NbN layer, and the highest $T_{c}$ in sample S-5, with a 140 nm thick NbN layer. This decline in $T_{c}$ can be attributed to the reduction in carrier density as the thickness of the NbN layer is decreased \cite{chockalingam2008superconducting, LICATA2022125962}. The experiments involving spin pumping and ISHE were conducted in a flip-chip geometry [Fig. \ref{Fig1}(a)] at a fixed frequency and variable field.\\
When a non-magnetic metal (NM) layer interfaces with magnetic materials (FM), such as Py, a  phenomenon known as spin pumping occurs \cite{tserkovnyak2002enhanced,tserkovnyak2002spin,brataas2002spin}. The precessing magnetic material transfers angular momentum from the FM layer to the FM/NM interface, resulting in an enlargement of the ferromagnetic resonance (FMR) linewidth within the magnetic layer. This magnetization oscillation at the FM/NM interface induces a spin imbalance in the NM layer, thereby generating a spin current within the NM. Subsequently, these spin-polarized electron currents are converted into conventional charge currents through the ISHE, providing a convenient means for electrically detecting spin-wave spin currents.\\
	The FMR  is one of the prominent techniques to determine the dynamic magnetic properties.  We curve-fitted the  FMR spectrum [Fig. \ref{Fig2}(a)]  using the equation (\ref{eq_FMR}) where $\Delta H$ and $ H_{r} $ are the full width at half maximum (FWHM) linewidth and resonance field of the FMR spectrum at a given excitation frequency, respectively \cite{maksymov2015broadband,martin2020temperature}.  $ A $ and $ B $ represent the amplitude of symmetric and asymmetric parts, respectively. 
	{\begin{equation}\label{eq_FMR}
			\dfrac{dP}{dH} = A          \frac{4\Delta H (H-H_{r})}{\Delta H ^{2}+4(H-H_{r})^{2}} + B \frac{[\Delta H ^{2}- 4 (H-H_{r})^{2})] }{\Delta H ^{2}+4(H-H_{r})^{2}} + offset  
		\end{equation}	
		We determined [using equation (\ref{eq_FMR})] essential parameters, namely, the resonance field $ H_{r} $ and the FWHM of the absorption peak $\Delta H$. By examining the frequency-dependent behavior of these fitting parameters, we calculated key dynamical properties of the sample, including the effective magnetization ($ M_{eff} $) using equation (\ref{eq_Kittel}),  damping constant ($\alpha$), and inhomogeneous broadening ($ \Delta H_{0} $) via equation (\ref{eq_damping}), where 	 $\lvert \gamma \rvert$ (= 1.76 x 10\textsuperscript{11} radT\textsuperscript{-1}s\textsuperscript{-1}) and $\mu_{o}$, are the gyromagnetic ratio and vacuum permeability, respectively.
		{\begin{equation}\label{eq_Kittel}
				f = \frac{\gamma\mu_{o}}{2\pi} \sqrt{H_{r} \left(H_{r} + M_{eff}\right)}       
			\end{equation}
			
			{\begin{equation}\label{eq_damping}
					\mu_{o}\Delta H = \frac{4\pi\alpha}{\gamma} f + \mu_{o}\Delta H_{0}  
				\end{equation}	
The effective magnetization values show only around 5\% increase from 300 to 20 K [Fig. \ref{Fig2}(b)]. A strong linear relationship between $f$ and $\Delta H$ was seen in all our samples between temperatures 300-20 K [Fig. \ref{Fig2}(c)].	 
All of the examined samples displayed pronounced non-monotonic temperature-dependent damping characteristics [Fig.  \ref{Fig2}(d)].
The contributions to the measured linewidth can be categorized into two distinct components: extrinsic effects, which originate from sample imperfections and inhomogeneities, and intrinsic effects, which arise from inherent interactions between the magnetic and non-magnetic modes that are intrinsic to the material. Notably, the residual linewidth in sample S-0 remains below 5 Oe, indicating a minimal impact from surface inhomogeneities \cite{jeon2019effect}. Moreover, it is important to highlight that two-magnon scattering cannot account for the observed reduction in linewidth as temperature decreases from 300 to 150 K [Fig.  \ref{Fig2}(d)]. The intrinsic damping, associated with classical eddy currents, known to increase proportionally with rising conductivity at lower temperatures, was found to be two orders of magnitude lower than the observed damping in the S-0 film.\\
Previous investigations into the temperature dependence of $\alpha$ in Py thin films have produced a wide range of intriguing findings. In certain instances, a slight decreasing trend \cite{counil2006temperature} in $\alpha$ has been observed, which has been attributed to reduced scattering at low temperatures. However, as the thickness of the film decreases, other reports have pointed to the emergence of a peak in the range of 30–40 K \cite{zhao2016experimental}. This peculiar behavior has been attributed to the influence of an additional dissipation term in thin films and multilayers, stemming from surface and interface effects \cite{barati2014gilbert}. Notably, these effects exhibit distinct temperature dependencies. Consequently, the damping behavior can be described as a combination of contributions from the bulk and the surface, with the observed cusp in $\alpha(T)$ being linked to a thermally induced spin reorientation of the surface magnetization \cite{zhao2016experimental,sierra2011interface}.

The noticeable increase in $\alpha$, resulting from the process of spin pumping into the NbN layer, is readily apparent. This observation points to the primary source of damping in Py/NbN bilayers, which is the loss of angular momentum during magnetization precession. At the resonance frequency, the pumped spin current reaches its maximum magnitude. To delve deeper into whether this observed rise in $\alpha$ can be attributed to the injection of spin current into the NbN layer, we conducted measurements of voltage as a function of the applied magnetic field.
We observed a significant dip in voltage at a magnetic field corresponding to the FMR  absorption [Fig. \ref{Fig3}(a),(c)]. Differentiating between the contributions of spin current and other factors is a complex and challenging endeavor, and various approaches have been proposed based on their distinct symmetry properties \cite{mosendz2010detection,bai2013universal,bai2013distinguishing}. To disentangle these contributions, we subjected the observed voltage data to fitting using Equation (\ref{eq_volt}), which allowed us to separate the components associated with the ISHE and spin rectification \cite{saitoh2006conversion}.

{\begin{equation}\label{eq_volt}
		V_{Data} = V_{Sym}\frac{\Gamma ^{2}}{\Gamma ^{2}+(H-H_{r})^{2}} + V_{Asym} \frac{-2\Gamma (H-H_{r})}{\Gamma ^{2}+(H-H_{r})^{2}}   
	\end{equation}
	
	where $ \Gamma  $ is the  half linewidth of the FMR line shape and H\textsubscript{r} is the resonance field. The first term  $V_{Sym}\frac{\Gamma ^{2}}{\Gamma ^{2}+(H-H_{r})^{2}} $ in equation (\ref{eq_volt}) is symmetric and has the Lorentzian form. The second term $ V_{Asym} \frac{-2\Gamma (H-H_{r})}{\Gamma ^{2}+(H-H_{r})^{2}} $ in equation (\ref{eq_volt}) is asymmetric with a dysonian dispersion line shape and has the contributions from the spin rectification effects such as anisotropic magnetoresistance  and the anomalous Hall effect  \cite{saitoh2006conversion,iguchi2017measurement} . The $ V_{Asym}$ changes sign as the magnetic field is swept through $H_{r}$. The symmetric voltage corresponds to the ISHE voltage. \\				
We direct the microwave current through the signal line of the coplanar waveguide  positioned at a height of 40 micrometers from the NbN layer [Fig. \ref{Fig1}(a)]. This spatial arrangement effectively minimizes the impact of microwave currents within the NbN layer. Furthermore, we utilize low radio-frequency  power excitation, resulting in a small cone angle for the magnetization precession. This can be observed from the linear frequency dependence of the absorption linewidth  [Fig. \ref{Fig2}(c) and Fig. \ref{Fig3}(e)]. As illustrated in Figure \ref{Fig3}(b), the spectral profile of the ISHE voltage signal exhibits remarkable symmetry, which can be accurately fitted using primarily a symmetric Lorentzian shape [Fig. \ref{Fig3}(b)]. This reaffirms that the contribution of spin rectification effects is indeed negligible within our experimental setup. When we examine the temperature dependence of the ISHE voltage, normalized by the symmetric component ($A$) of Equation (\ref{eq_FMR}), and the corresponding resistance value at a given temperature, we observe an increase in these values as the temperature is lowered [Fig. \ref{Fig3}(f)].\\
The spin mixing conductance g\textsubscript{sm} is an inter-facial parameter that quantifies amount of spins flown from the magnetic material under FMR to the NbN. The g\textsubscript{sm} values in NbN capped samples were calculated using equation (\ref{eq_gsm}) \cite{ando2010inverse,wang2016surface}.
{\begin{equation}\label{eq_gsm}
		g_{sm} = \frac{4\pi M_{S}  t_{py}}{g  \mu_{B}} (\alpha_{Py/NbN}  -  \alpha_{Py} )   
	\end{equation}
 $t_{py}$, $\alpha_{Py/NbN}$, $\alpha_{Py}$, $ g $, and $\mu_{B}$ are the    thickness of the Py film,  damping of NbN capped Py samples, damping of Py film, Lande-g factor, and Bohr magneton, respectively.  
	We observed that the  $g_{sm}$ reaches its maximum value for sample S-5 [Fig. \ref{Fig4}(a)]. Specifically, it increased from 2.37 x 10\textsuperscript{18} 1/m\textsuperscript{2} at 300 K to 6.48 x 10\textsuperscript{19} 1/m\textsuperscript{2} at T = 30 K. Conversely, the lowest $g_{sm}$ was found for sample S-4. Moreover, the difference between the $g_{sm}$ values was relatively small at T = 300 K but exhibited substantial deviations at lower temperatures. This observation underscores the heightened sensitivity of interfacial coupling between NbN and Py at lower temperatures, as well as the increased dependence on bilayer growth conditions at reduced temperatures. In summary, the $g_{sm}$ values displayed a consistent trend of increasing as the temperature of the samples decreased.\\
	The magnitude of the spin current injected from the Py to the NbN layer [Fig. \ref{Fig4}(b)] can be calculated from the spin current density formulation   \cite{tserkovnyak2005nonlocal,ando2010inverse}  and  can be expressed as 		
		{\begin{equation}\label{eq_Js}
				\lvert J_{s} \rvert=\left( \frac{2e g_{sm }}{8\pi } \right)         \left( \frac{\mu_{o}h_{rf}\gamma}{\alpha_{eff} } \right)^{2} \left[ \frac{\mu_{o}M_{s}\gamma  +\sqrt{ (\mu_{o}M_{s}\gamma)^2 +16(\pi f)^2 }}{(\mu_{o}M_{s}\gamma)^2 +16(\pi f)^2}                \right]
			\end{equation}

			 $h_{rf}$ is the radio frequency field due to signal line of the waveguide, which for a 200 $\mu$m signal line width ($ w $) is 0.9 Oe \cite{arena2009compact}.
			The spin current density [Fig. \ref{Fig4}(c)] described in equation (\ref{eq_Js}) is converted into an electromotive force $V_{ISHE}$ due to the ISHE in the NbN layer induced by the spin pumping as per the following relation:
			
			{\begin{equation}\label{eq_ISHE}
					V_{ISHE} = \lvert J_{s} \rvert\frac{w\Theta_{SH}\lambda_{NbN} tanh\left(\frac{d_{p}}{2\lambda_{N}}\right)} {  d_{NbN}\sigma_{NbN}+d_{Py}\sigma_{Py} }   
				\end{equation}	
				The  $d_{NbN}$, $\sigma_{NbN}$,  $ d_{Py} $, $ \sigma_{Py} $,  $ \lambda_{NbN} $ (= 14 nm for NbN\cite{roy2017estimating}), and $ \Theta_{SH} $  are the thickness of NbN, conductivity of NbN, thickness of Py, conductivity of Py,  spin diffusion length of NbN layer, and the spin Hall angle (representing the efficiency of spin-to-charge conversion of NbN), respectively.	
Our spin pumping experiments have revealed a  difference in the $ \Theta_{SH} $ compared to prior quantifications by Wakamura et al., who employed lateral spin valve samples ($\Theta_{SH} \approx $ -0.9 x 10\textsuperscript{-2} at 20 K) \cite{wakamura2015quasiparticle}, and Rogdakis et al., who conducted their measurements using spin pumping methods \cite{rogdakis2019spin} found $\Theta_{SH} \approx $ -1.1 x 10\textsuperscript{-2} at room temperature. Specifically, our observations indicate that for sample S-5, $\Theta_{SH}$ is three times higher. This notable variation in $\Theta_{SH}$ values between studies within the temperature range of 20-300 K might be attributed to several factors. Firstly, our experimental approach involved the in-situ growth of Py/NbN bilayer samples, without subsequent microstructuring, which distinguishes it from the methods employed by Rogdakis et al. \cite{rogdakis2019spin} and Wakamura et al. \cite{wakamura2015quasiparticle}. The  growth conditions (e.g., lower base pressure of $ 10^{-9} $ mbar,  the density of scattering impurities and carrier densities \cite{chockalingam2008superconducting}, etc.)  and interfaces in our samples could potentially lead to distinct $\Theta_{SH}$ values. Secondly, it's crucial to recognize that the determination of $\Theta_{SH}$ for the same material can exhibit notable variations when conducted by different research groups. This phenomenon has been observed in various materials, including Pt \cite{liu2011review} and certain topological insulators \cite{mellnik2014spin,fan2014magnetization,deorani2014observation}. These disparities often arise from differences in sample quality, which can be influenced by growth conditions, thereby impacting $\Theta_{SH}$ through extrinsic spin Hall mechanisms. Our ISHE study on YIG(90 nm)/Pt(15 nm) show the sign of $V_{ISHE}$ is opposte to that for $Py/NbN$ layers, indicating $\Theta_{SH}$ is of negative sign for $Py/NbN$ layers, which  is in agreement with  Wakamura et al. \cite{wakamura2015quasiparticle} and Rogdakis et al. \cite{rogdakis2019spin}	\\
Cooper pairs in superconductors, like NbN, are characterized by spin singlet states, and as such, they are unable to carry a spin current. In singlet superconductors, spin current can only be transported by quasiparticle excitations \cite{morten2008proximity}. However, the low-energy density of quasiparticle states is suppressed by superconducting correlations, resulting in enhanced spin transport resistivity in the superconducting state. This leads to reduced spin injection from the ferromagnet compared to the normal state, which manifests as reduced damping in the magnetization dynamics. The temperature dependence of the FMR linewidth, both near and below the critical temperature, provides valuable insights into superconducting correlations in magnetic heterostructures and spin-flip processes. These insights have potential implications across various areas of mesoscopic physics \cite{morten2008proximity}.
The Py/NbN bilayers exhibit a significant variation in their  $T_{c}$, which depends on the thickness of the NbN layers. For instance, sample S-5 (with a NbN thickness of 140 nm) has a $T_{c}$ of 14.5 K, while the lowest $T_{c}$ of 10 K was observed in sample S-1 (with a NbN thickness of 20 nm) [Fig. \ref{Fig1}(b)]. Consequently, substantial variations in damping were observed both above and below the $T_{c}$.
Upon cooling the samples below their respective $T_{c}$, distinctive characteristics in ISHE voltage, FMR amplitude [Fig. \ref{Fig3}(b), \ref{Fig5}(a) and (b)], linewidth [Fig. \ref{Fig5}(c)], and resonance field [Fig. \ref{Fig5}(d)] emerge. The ISHE voltage drops below the noise level of the instrument,  [Fig. \ref{Fig3}(d)], while the FMR linewidth increases in value, particularly in samples with a thickness greater than 60 nm [Fig. \ref{Fig5}(c)]. This increase in linewidth contradicts the expected decrease in linewidth due to the decay in quasiparticle density below the $T_{c}$.  This anomalous behavior may also be attributed to the 15 nm thickness of the Py film in our samples, which is more than twice the exchange length (approximately 6 nm) for Py. Consequently, a slight misalignment of magnetization near the interface with NbN in such thick Py films may lead to a higher linewidth below $T_{c}$.
Our findings have the potential to stimulate further experimental and theoretical studies in these bilayer films. One avenue for exploration could involve reducing the thickness of the Py layer below the exchange length in such systems. Additionally, introducing an insulating layer between NbN and Py could help us understand whether the increased linewidth below $T_{c}$ persists or not. This is particularly relevant in the realm of magnon-photon coupling, where a relatively thick (30 nm) Py layer is often separated from a (200 nm) thick NbN layer by an insulating layer \cite{li2019strong}.\\
In conclusion, our investigation into broadband spin pumping via the ISHE in Py/NbN bilayer films has revealed robust temperature-dependent behavior in $\alpha$, $g_{sm}$, $ J_{s} $, and $ \Theta_{SH} $ parameters. Notably, our findings have unveiled significantly elevated $\Theta_{SH}$ values, which hold the promise of promoting the utilization of NbN in superconductivity-based quantum devices. Furthermore, it is worth highlighting the increased linewidth observed below the $T_c$, suggesting the need for further exploration through both theoretical and experimental approaches. 
\\
We acknowledges support from the  National Science Center in Poland under the grant number UMO-2020/38/E/ST3/00578 and Foundation for Polish Science through the IRA Programme financed by EU within SG OP Programme.
\\
\bibliography{Bibliography3}
\newpage
 \begin{figure}
	\includegraphics[width=0.5\textwidth]{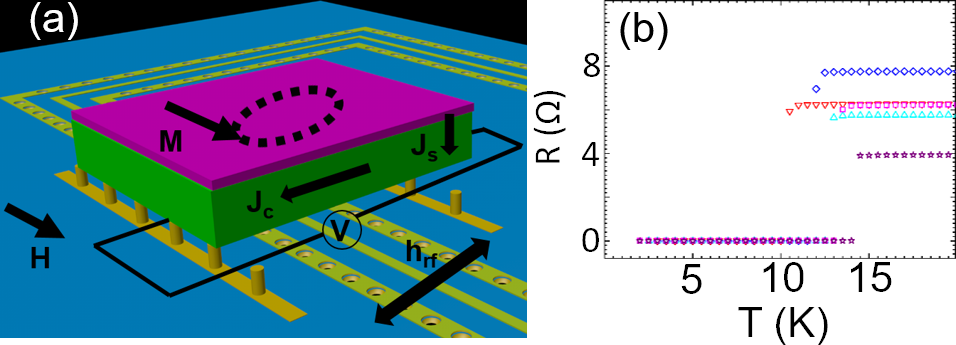}
	\begin{flushleft}
		\caption{ (a) Schematic of the spin pumping and ISHE setup. The magenta and green colors show Py and NbN layers, respectively. (b) Resistance vs. temperature for samples S-1 to S-5 in red, blue, cyan, magenta and purple-colored symbols, respectively. 
		}\label{Fig1}
	\end{flushleft}
\end{figure}				
\begin{figure}
	\includegraphics[width=0.5\textwidth]{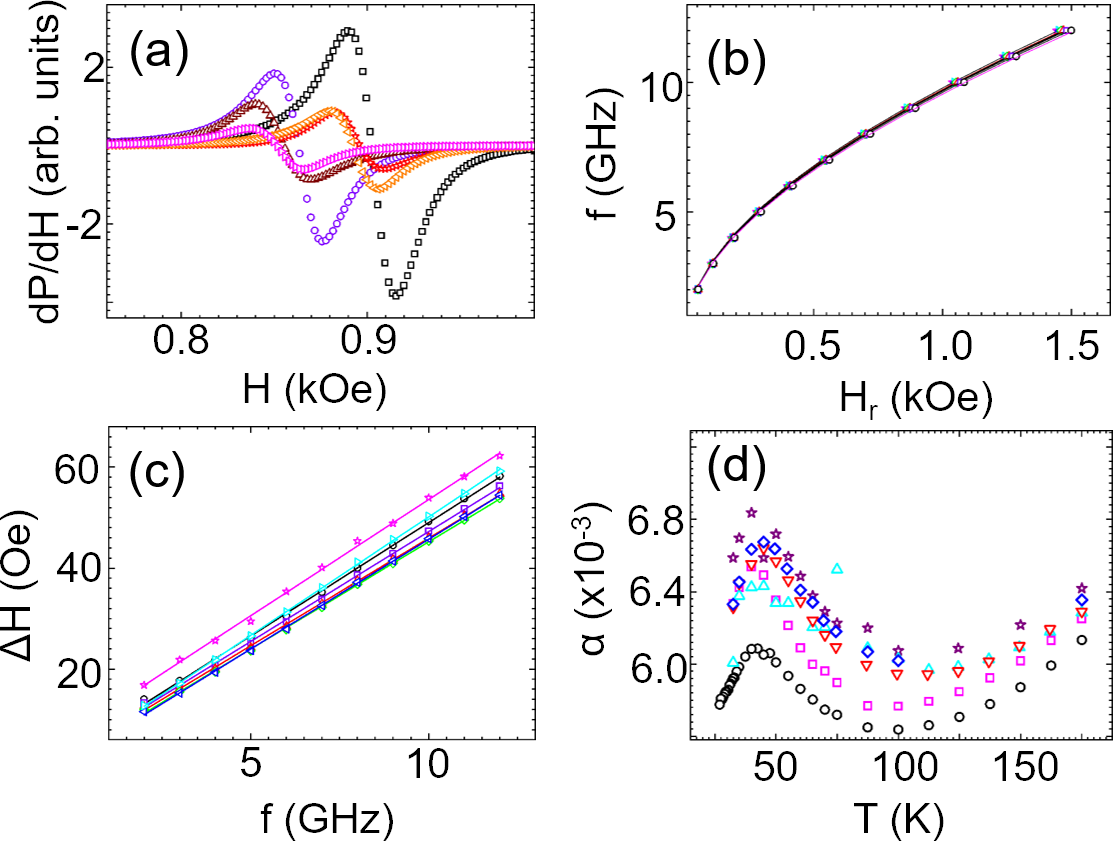}
	\begin{flushleft}
		\caption{ (a) The black (purple), orange (magenta), and red (wine) colors represent FMR spectra for samples S-0, S-1, and S-5 with T = 300 K (20 K), respectively. (b)
			Frequency vs. resonance field and (c) FWHM  vs. excitation frequency for sample S-1 at different temperatures. The symbols and lines indicate  data and fit, respectively. The black, purple, red, green, blue, cyan, and magenta colors represent T = 300, 250, 200, 150, 100, 50, and 20 K, respectively. (d) Temperature dependence of damping for all the samples. The black, red, blue, cyan, magenta, and  purple-colored symbols denote the samples S-0 to S-5, respectively.	 
		}\label{Fig2}
	\end{flushleft}
\end{figure}
\begin{figure}
	\includegraphics[width=0.5\textwidth]{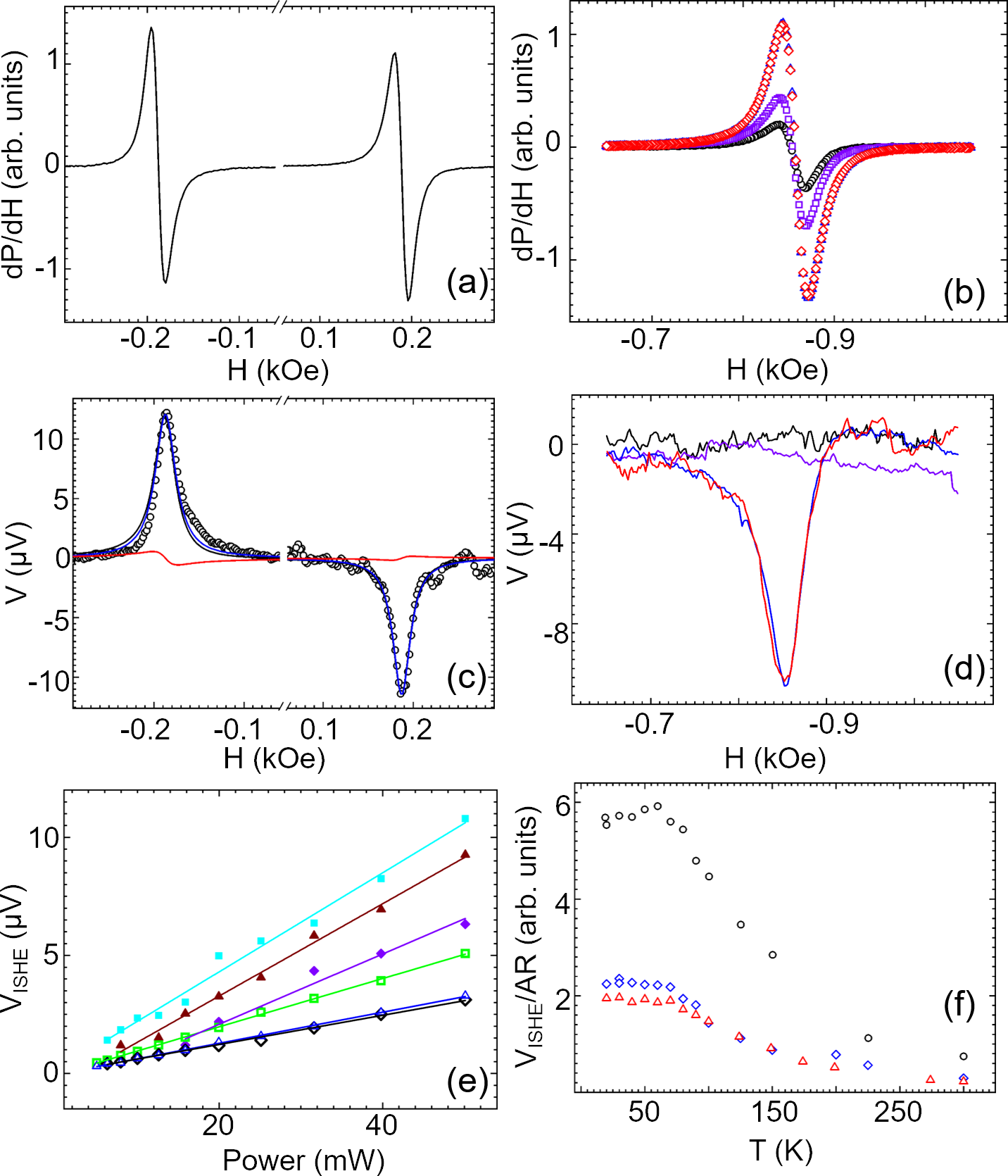}
	\begin{flushleft}
		\caption{ (a) The FMR spectra and (c) corresponding voltage  
			for positive and negative field sweeps at f = 9 GHz and T =20 K for the sample S-1. The black, red and blue colors in (c) indicate data, symmetric ISHE fit and asymmetric spin rectification fit, respectively. (b) The FMR spectra and (d) corresponding voltage at 9 GHz for the sample S-1. The blue, red, violet and black colors represent data at temperatures 11, 10, 9, and 8 K, respectively.  (e) Power dependence of the peak of ISHE   for sample S-1 at two different temperatures. The symbols represent data and solid lines indicate corresponding linear fit. The cyan, violet, and wine colors show frequencies 4, 7, and 9 GHz at T = 20 K, respectively, whereas the green, black and blue colors demonstrate results for excitation frequencies 4, 7, and 9 GHz at T = 300 K. (f) Temperature dependence of the normalized ISHE voltage with respect to FMR amplitude and resistance for sample S-1. The black, blue and red colors represent f = 4 ,7, and 9 GHz, respectively.   
		}\label{Fig3}
	\end{flushleft}
\end{figure}
\begin{figure}
	\includegraphics[width=0.5\textwidth]{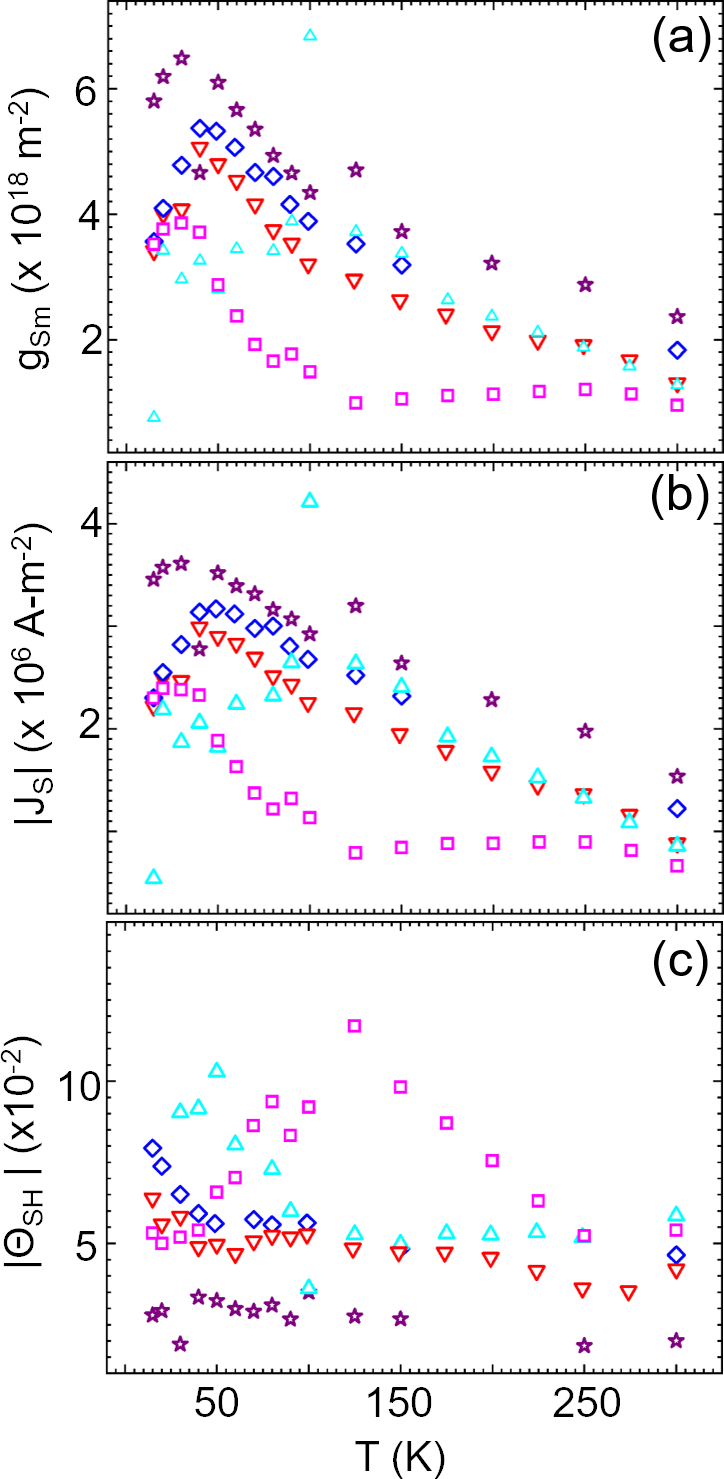}
	\begin{flushleft}
		\caption{ The temperature dependence of (a) spin mixing conductance, (b) spin current density and (c) spin Hall angle values  for studied samples. The red, blue, cyan, magenta, and purple-colored symbols represent samples S-1, S-2, S-3, S-4, and S-5, respectively.  
		}\label{Fig4}
	\end{flushleft}
\end{figure}

\begin{figure}
	\includegraphics[width=0.5\textwidth]{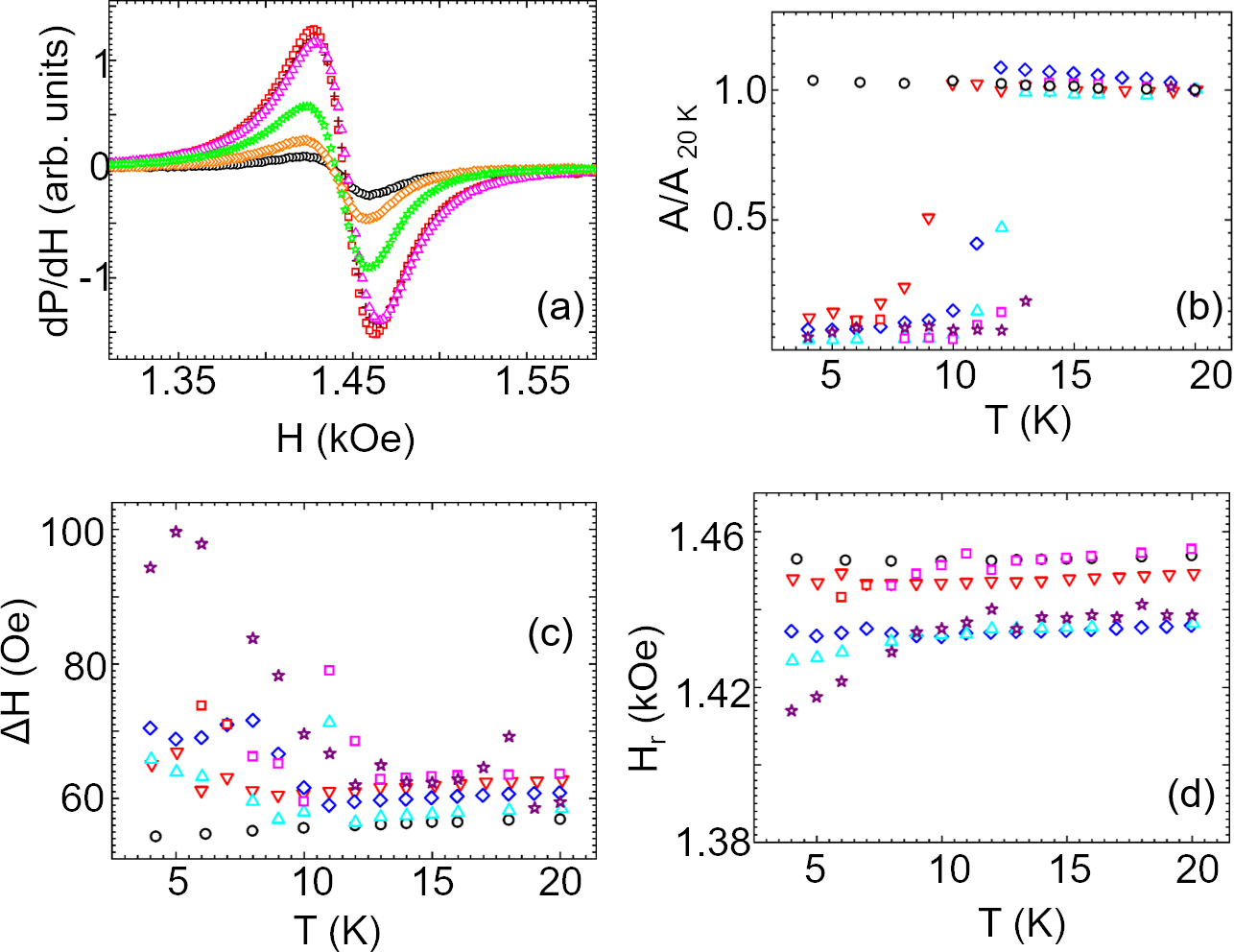}
	\begin{flushleft}
		\caption{(a) The temperature dependence of FMR spectra at f = 12 GHz for sample S-1.  Black, orange, green, red, wine, and magenta colored symbols represent FMR spectra for 5, 8, 9, 10, 15, and 20 K, respectively. (b) Symmetric peak amplitude for studied samples at f = 12 GHz. We normalized the amplitudes  with respect to the value at T = 20 K for a given sample. (c) FWHM, and (d)  resonance field values at f = 12 GHz for studied samples. The black, red, blue, cyan, magenta and purple colored symbols in (b)-(d) represent values for samples S-0, S-1, S-2, S-3, S-4, and S-5, respectively. 		}\label{Fig5}
	\end{flushleft}
\end{figure}
\end{document}